\documentclass[conference]{IEEEtran}
\ifCLASSINFOpdf
\else
\fi

\usepackage{stfloats}
\usepackage{balance}  
\usepackage{graphics} 
\usepackage{url}      
\usepackage{times}    
\usepackage[pdftex]{hyperref}
\usepackage{blindtext, graphicx}
\usepackage{url} 
\usepackage{color}
\hypersetup{draft}

\begin{document}
%
\title{A Comparison of Common Users across Instagram and Ask.fm to Better Understand Cyberbullying}

\author{\IEEEauthorblockN{Homa Hosseinmardi \\ Shaosong Li, Zhili Yang and Qin Lv}
\IEEEauthorblockA{Department of Computer Science\\
University of Colorado at Boulder\\
Boulder, Colorado, USA\\
Email: homa.hosseinmardi@colorado.edu}
\and
\IEEEauthorblockN{Rahat Ibn Rafiq \\ Richard Han and Shivakant Mishra}
\IEEEauthorblockA{Department of Computer Science\\
University of Colorado at Boulder\\
Boulder, Colorado, USA\\
Email: rahat.rafiq@colorado.edu}}


%


\maketitle

\begin{abstract}
This paper examines users who are common to two popular online social networks, Instagram and Ask.fm, that are often used for cyberbullying.  An analysis of the negativity and positivity of word usage in posts by common users of these two social networks is performed.  These results are normalized in comparison to a sample of typical users in both networks.  We also examine the posting activity of common user profiles and consider its correlation with negativity.  
Within the Ask.fm social network, which allows anonymous posts, the relationship between anonymity and negativity is further explored.
\end{abstract}

\section{Introduction}

The rise of online social networks (OSNs) has resulted in a significant increase in
cyberbullying activities, especially among teenagers. Indeed, 
cyberbullying has become a major societal problem. For example, Natasha MacBryde, a 15-year old girl, committed suicide after receiving bullying comments from anonymous users on the Formspring social network and being called ``slut'' by her high school friends~\cite{Natasha}. Hannah Smith, a 14-year old girl, killed herself after being cyberbullied on the Ask.fm social network \cite{hannah}. There is an urgent need for detecting incidents of cyberbullying in OSNs, so that appropriate help can be provided to the cyberbullying victims in a timely manner and bullies can be identified and stopped. 

Detection of cyberbullying in OSNs is challenging for a number of  
reasons. First, the sheer volume of postings in OSNs renders 
any manual attempts to detect cyberbullying completely infeasible.
Second, it is not at all clear what constitutes cyberbullying as it depends
on a number contextual factors. Finally, there are a multitude of OSNs that differ from one another in a variety of ways, such as the type
of user postings (text, images, videos, etc.), ability to post messages
anonymously, methods of responding to user posts (liking individual postings
vs. a single response to a group of postings), ability to make friends vs. 
following a user, whether the primary method of posting is via a mobile
phone, etc. Therefore, an 
effective solution for automatically detecting cyberbullying must account for variations among different OSNs. 

In our previous work~\cite{ASONAM}, we
analyzed negative user behaviors at Ask.fm, a popular
OSN that has led to many cases of cyberbullying, some resulting in suicide. 
We examined the occurrence of negative words in users' 
question+answer profiles along with the “likes” of these questions+answers. 
Our results indicate that negativity exhibited in users' online behaviors, 
such as the type of words used in postings, type of postings liked, and level 
of OSN activities,  can potentially help identify cyberbullying. 

In this work, we tackle the challenge of understanding how negative user behavior
varies across different OSNs, in particular the two OSNs Ask.fm and Instagram. Both OSNs are very 
popular among teenagers and also rank among the top sites for cyberbullying~\cite{annual}. These two networks differ from each another in some significant ways. Ask.fm is a semi-anonymous network in which user profiles 
are public, but postings to these profiles by non-owner users are anonymous 
by default. On the other hand, Instagram makes the identities of user 
postings public. Furthermore, users in Ask.fm can post 
questions and profile owners can provide answers to those questions
individually, while users in Instagram can share and comment on 
media objects such as images.

To understand how these differences impact user behavior, we 
collected comprehensive data on four types of users: 
(1) {\em normal Instagram users}, 
(2) {\em normal Ask.fm users}, 
(3) {\em common Instagram users} who are also Ask.fm users, and 
(4) {\em common Ask.fm users} who are also Instagram users. 
We then compare the posting behaviors of normal and common users 
in each social network, as well as the posting behaviors of common 
users across both networks.
Together, these comparisons provide a deeper understanding
of user behavior both within each OSN as well as across two different OSNs. 

Our analysis in this work has revealed the following five important 
findings. First, we show that more negativity is exhibited at Ask.fm
than at Instagram. Second, we show that there is no significant difference
in the positivity exhibited by the users in the two social networks.
Third, we show that within each social network, common and normal users tend to behave similarly
in terms of their positivity and negativity. Fourth, we show that anonymity tends to result in 
lower negativity. This finding is quite counter-intuitive and we provide some
possible reasons for it. Finally, we provide a detailed correlation  
analysis among different user behaviors and activities in the two networks.

\section{Related Work}

There have been a variety of studies in the fields of education and psychology on the prevalence and impact of cyberbullying~\cite{students-cyberbullying, youth-interaction, defining-cyberbullying}. In this work, we focus on 
understanding how OSNs are being used to enable cyberbullying. 
Prior works on the analysis and detection of cyberbullying in OSNs have largely focused on social networks such as YouTube, Formspring, MySpace, Instagram and Twitter~\cite{Dinakar_modelingthe2, dadvar, Twitter}. Dinakar {\em et al.} investigated both explicit and implicit cyberbullying by analyzing negative text comments on YouTube and Formspring profiles~\cite{Dinakar_modelingthe2}. Dadvar {\em et al.} investigated how combining text analysis with MySpace user profile information such as gender can improve the accuracy of cyberbullying detection in OSNs~\cite{dadvar} . Sanchez and Kumar proposed using a Naive Bayes classifier to find inappropriate words in Twitter text data for bullying detection~\cite{Twitter}. They tracked  potential bullies, their followers, and the victims.
All these works focused on text-based analysis of negative words, and did not explore social network relationships in their investigation of cyberbullying.
Homan {\em et al.} studied the social structure of LGBT
youths with depression in the TrevorSpace social network~\cite{depressionTre}.
In this paper, we not only compare the language model of the two
social networks in terms of negative user behavior, but also 
investigate its relations to social network features and user activities.

Both Ask.fm and Instagram have been used for cyberbullying. 
According to a recent survey~\cite{annual}, Ask.fm
ranks as the fourth worst site in terms of the percentage of young users bullied.  Indeed,  a survey from 13-16 year old students of a British school showed a higher level of  abuse in Ask.fm than that of  Facebook and Twitter~\cite{formspring}.  Ashktorab {\em et al.} studied Ask.fm questions+answers and introduced a tool, iAnon, which tries to help victims of cyberbullying by sending them positive messages~\cite{iAnon}.  
Similarly, the Cyberbullying Research Center reports different ways by which cyberbullying occurs on Instagram~\cite{cyberinsta}. These include posting a malicious or embarrassing photo of a target for all the followers to see or posting cruel comments under a photo that someone posts.  
Other research on Instagram did not explore cyberbullying on the network, e.g.,  \cite{Weilenmann} investigated users' photo sharing experience in a museum while 
\cite{Insta} only looked at temporal photo sharing behavior of Instagram users, and \cite{whatinsta} categorized Instagram images into 8 groups.
Previously \cite{pintweet} has compared Pintrest (image-based social network) and Twitter (text-based social network) to understand the posting behavior across multiple sites and to find out the dynamics of sharing information, but not in the context of cyberbullying. To the best of our knowledge, this paper is the first work on cyberbullying that compares negative (and positive) user behavior across both a semi-anonymous social network (Ask.fm) and a non-anonymous mobile social network (Instagram), using both normal users and common users of the two social networks. 

Recently, \cite{anon13} has considered the impact of anonymity in TechCrunch.com, a technical social news site, where users are allowed to comment anonymously. On another work, the effect of anonymity on Facebook video sharing concluded that familiarity with the content of the video is a contributing factor in sharing behavior \cite{anon2}. Unfamiliarity and controversiality of contents has led to more anonymous sharing.

\section{Methodology} 
\label{sec:methodology} 

To effectively detect cyberbullying in OSNs, we need to 
understand if and how cyberbullying is affected by OSNs with different 
features. We chose to study Instagram and Ask.fm in this work, because both 
networks are very popular among teenagers and have increasingly been used 
for cyberbullying. More importantly, we have noticed that some Ask.fm users mentioned their Instagram IDs in their profiles, thus allowing for direct
comparison of user behaviors across these two social networks. 

In this work, we consider four different types of users: 
\begin{itemize}
\item {\bf Normal Instagram users}: A sample set of Instagram users with public profiles 
\item {\bf Normal Ask.fm users}: A sample set of Ask.fm users
\item {\bf Common Instagram users}  A sample set of Ask.fm users, for whom we also have collected their Instagram profile data 
\item {\bf Common Ask.fm users} A sample set of Instagram users, for whom we also have collected their Ask.fm profile data
\end{itemize} 

We are interested in answering the following questions: 
\begin{enumerate} 
\item Do common and normal Instagram users differ in their posting behaviors? 
\item Do common and normal Ask.fm users differ in their posting behaviors? 
\item Do common Instagram users and common Ask.fm users differ in their posting behaviors? 
\end{enumerate} 

We consider profanity words usage as a potential indicator of cyberbullying. Therefore we are mostly interested in the negative and positive words usage and the difference across Ask.fm and Instagram among common users. For this purpose we are using a dictionary of 1500 negative words (obtained from \cite{NegativeWordsList}) and 1500 positive words (obtained from \cite{posword}). A subset of abbreviations and symbols based on non-standard writing of negative words (e.g. wtf or a55) have been added to the list of words. As a preprocessing step, we applied stemming and removal of punctuation such as (!, ?, ", etc.) at the end of words to each comment. Then we calculate the percentage of posts by profile owners and friends/non-owners that include at least one of the words in our dictionaries and consider it as a measure of \textit{negativity} and \textit{positivity} of a user profile. While a list-based approach to detecting negativity may produce false positives, we feel it is a good starting point to provide insight into the usage of profanity language by social network users.

\subsection{Instagram and Ask.fm Features}

In Ask.fm, each user maintains his or her profile and other users can ask
questions by posting a question on that profile.  Ask.fm is a semi-anonymous
network in which the identities of the profile owners are always known but 
by default the identities of the users asking questions are not revealed. 
However, users asking questions have the option to reveal their identities.  
In the Ask.fm data that
we analyzed, we noticed that on average, about 27\% of the posts in users' 
profiles are non-anonymous. Profile owners may answer
some of the questions posted on their profile. Their web pages show only the
questions they choose to answer along with their answers. Users have the
option of liking a particular conversation consisting of a question and
answer pair which can not be done anonymously. They also
have the option to follow another user as well as send private, public, or
anonymous gifts to the other users.

While Ask.fm is a web and mobile-based 
social network that focuses primarily on textual content, Instagram
is primarily a mobile social network whose main feature is allowing users to take pictures using their
smartphones.  These pictures can then be digitally filtered and posted as media on their
accounts, and shared through other social networking sites such as Facebook,
Flickr, Twitter, and so on. Users can follow other users after their follow requests have been accepted. A user following another user can see the  media posts of that user. The default access privilege in Instagram is public, which means 
everyone can see the media objects posted by a user. However, users have
the option to change the access privilege to private, so that only the
people who are following them can see their posts. In the Instagram data
that we analyzed, we noticed that 39\% of users chose to keep their
access privilege private. Users in Instagram can like a media post, comment
on a media post, and can also tag other users while commenting. Unlike 
Ask.fm, Instagram is a completely non-anonymous network since the names 
and profiles of users who like, follow, or comment are visible to everyone.

\subsection{Data Collection}

To compare user posting behaviors across Instagram and Ask.fm, we collected a comprehensive set of user data from each network. For Instagram users, starting from a seed node, 41K user ids were gathered with a snowball sampling method. Among these Instagram ids, there were both public and private profiles, of which only 61\% of them have public profiles, or about 25K public profiles. 
These 25K public user profiles are used as our {\bf normal Instagram users} data.  For each public Instagram user, the collected profile data includes the media objects, the comments, the user id of each person followed by the user, 
 the user ids of those who follow the user,  and similarly for comments and likes of all shared media objects.  For Ask.fm, starting from a seed node and again 
using snowball sample, 24K complete Ask.fm user profiles were collected and used as the {\bf normal Ask.fm users} data. For each Ask.fm user, we collected the profile information, complete list of questions, their answers and authors (if not anonymized) and user ids of people who have liked each question+answer pair. 

After collecting Ask.fm user information we observed some users have revealed their Instagram ids in their profile information. Selecting those users that have both their Ask.fm and Instagram ids enables us to better compare two networks than comparing aggregates/averages of normal but non-common users.  However, the incidence of such common users is relatively low, so we needed to collect a large number of Ask.fm profiles to find a sufficient number of common profiles.  For this purpose, starting from a seed node, we collected 1M users' information via snowball sampling. Only 4\% of the users have mentioned their Instagram id in their profile information, thereby furnishing around 40K users with ids for both social networks.  Only about 24K of these profiles were public.  From this, we collected complete profile content for an 8K subset of these common users from each of the two social networks, forming our 
{\bf common Instagram users} data and {\bf common Ask.fm users} data.
\section{Analysis of Instagram User Behavior}

\begin{figure}[!h]
\centering
\includegraphics[width=1\columnwidth]{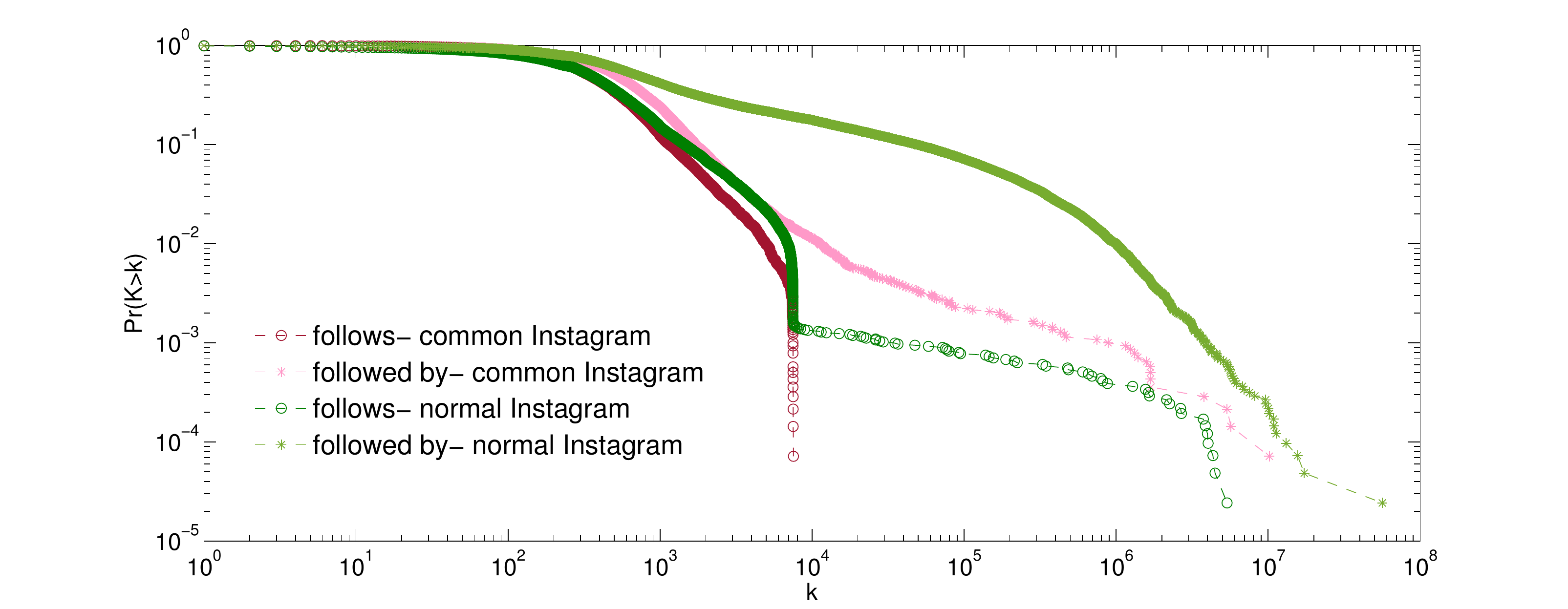}
\caption{Complementary cumulative distribution functions (CCDFs) of {\em follows} and {\em followed by} of normal and common Instagram users.}
\label{flld_flls_insta}
\end{figure}

In this section we examine the distributions of different types of activities of 
normal and common Instagram users. Figure \ref{flld_flls_insta} shows the complementary cumulative distribution
function (CCDF) of the number of
users a user follows ({\it follows}) and the number of users a user is
followed by ({\it followed by}) in the Instagram social network. These CCDFs
are shown separately for normal Instagram users and common Instagram users. Note 
that only those Instagram users who make their profiles public are included
in this study. It can be seen that for both common and normal users, 
there is a large gap between the number of followed bys and follows.
The main reason is that Instagram imposes 7,500 as the maximum
number of users a user can follow. This was done to limit spam.
There is no such limit on the number of users a user can be followed by.
Also, we note that the follows behavior is similar for both common and normal
users, except that there is a long tail in ``normal follows''.  This is due to the fact that this limit was
imposed in January 2013, and the users who were following more than 7,500
users before then continue to do so. Our examination suggests that most of these accounts in this long tail
belong to campaigns and advertisements as well as
aspiring photographers, models, businesses and  people trying to attract 
attention on the popular page.  Comparing the followed by behavior, we observe that there is a large gap between common and normal users' curves, but  the two curves
eventually converge, and  both have long tails.  We believe that
these long tails (up to $10^7$) are due to the presence of celebrities who tend to be followed by
a significantly larger number of users.  We also have found a small number of celebrities among our
common users, which lead to the convergence of the two curves.
In terms of the large gap in followed bys between normal and common Instagram users, it probably comes from the nature of the Instagram social network which is a photo sharing website. Users followed by more than $10^3-10^5$ people are less likely to appear in the set of common users. Looking at normal users'  profiles, we see users ranging from small music bands, local celebrities or fashions, popular nonprofessional photographers. These users seem to be interested in the photo sharing nature of Instagram and tend not to have profiles in Ask.fm. At the tail we have celebrities, campaigns and advertisements which open accounts at all social networks and that is why the two CCDFs converge.

\begin{figure}[!h]
\centering
\includegraphics[width=1\columnwidth]{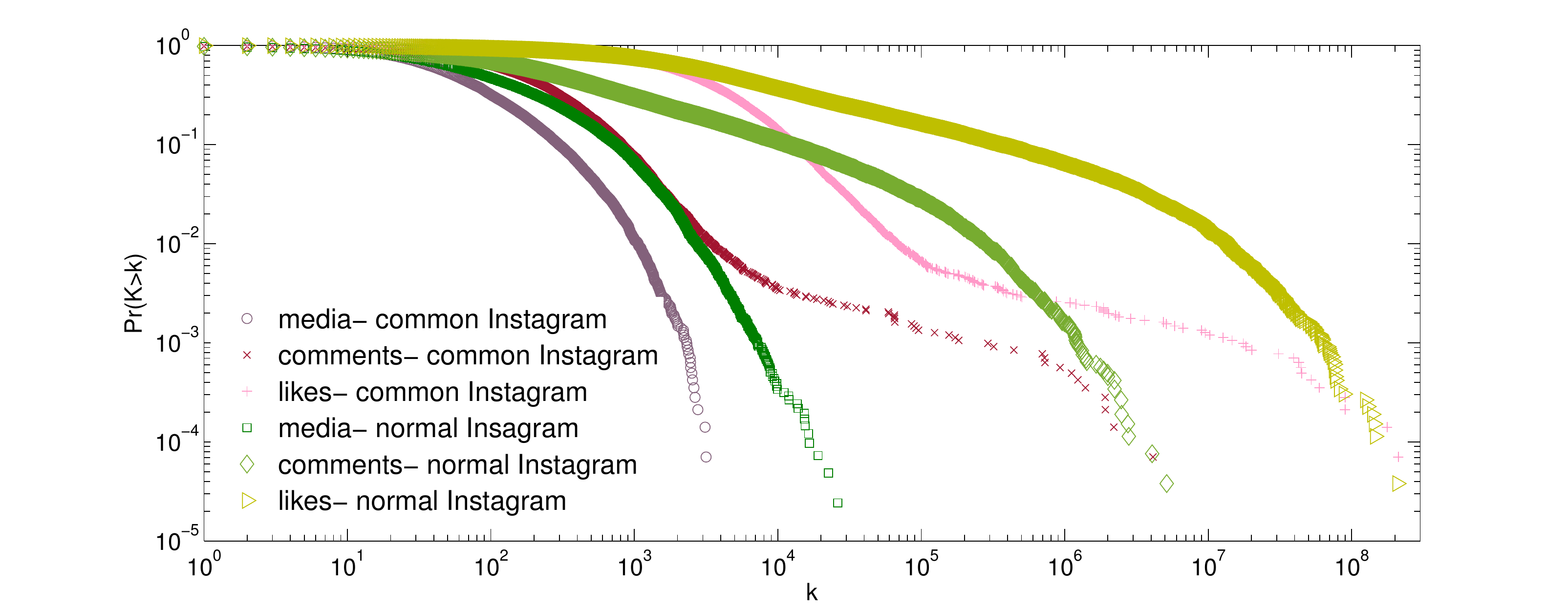}
\caption{CCDFs of the number of shared pictures, number of likes received, and number of comments, of normal and common  Instagram users.}
\label{media_likes_comments_insta}
\end{figure}

Next, we analyze the following Instagram user activities: 
the number of media objects shared by a user, 
the number of likes received on the user's media objects, and 
the number of comments (by others and the owner) on a user's media objects. 
Figure \ref{media_likes_comments_insta} shows the CCDFs of these three user
activities for normal and common Instagram users. We notice that for all these three
user activities, normal users are generally more active than common users.
We again observe big gaps between common and normal likes and comments. As will be 
shown later in Section \ref{sec:commn}, there exist fairly high correlations among followed by, likes, and comments.  We can see  that their CCDFs have 
the same pattern.

\begin{figure}[!h]
\centering
\includegraphics[width=1\columnwidth]{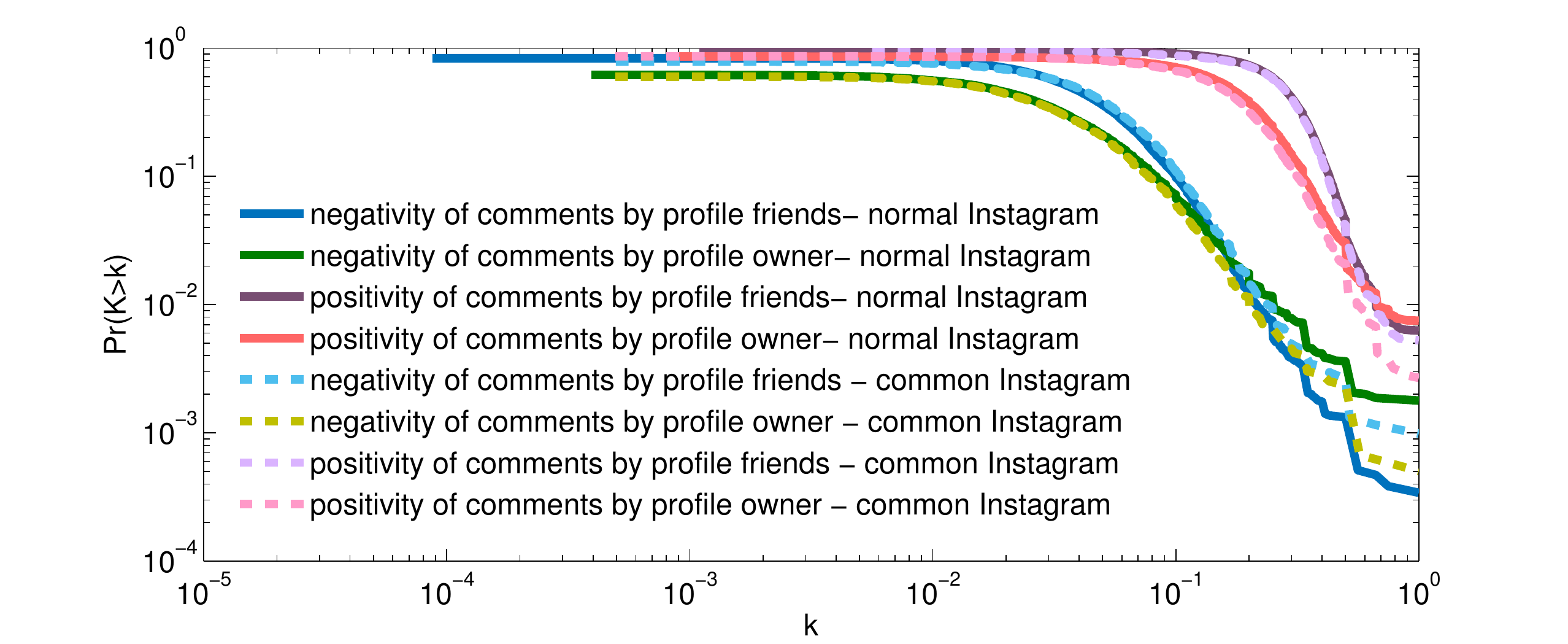}
\caption{CCDFs of the percentage of negative and positive posts by normal and common public Instagram users. Red and purple show positivity and blue and green show negativity.}
\label{insta_total}
\end{figure}

Finally, we analyze the user behavior in terms of negativity and positivity
in Instagram. To gain  deeper insights, we divide all user posts into two groups, posts by profile owners and posts by other non-owner users, whom we will loosely term ``friends'' in the remainder of this paper though some of these other users may not be truly friendly.
Figure \ref{insta_total} shows the CCDFs of negativity and positivity of
comments posted by profile owners and their friends, and again separated for
normal and common Instagram users. In this figure, dashed lines are the CCDFs for common users  and solid lines are for normal users. The first
observation we make here is that
positivity is always greater than negativity for both profile owners and
their friends as well as for normal and common users.
\emph{
The second important observation we make is that the user behavior is
quite similar for normal and common users both for negative and positive user
behaviors. 
}

%

\section{Analysis of Ask.fm User Behavior}

\begin{figure}[!h]
\centering
\includegraphics[width=1\columnwidth]{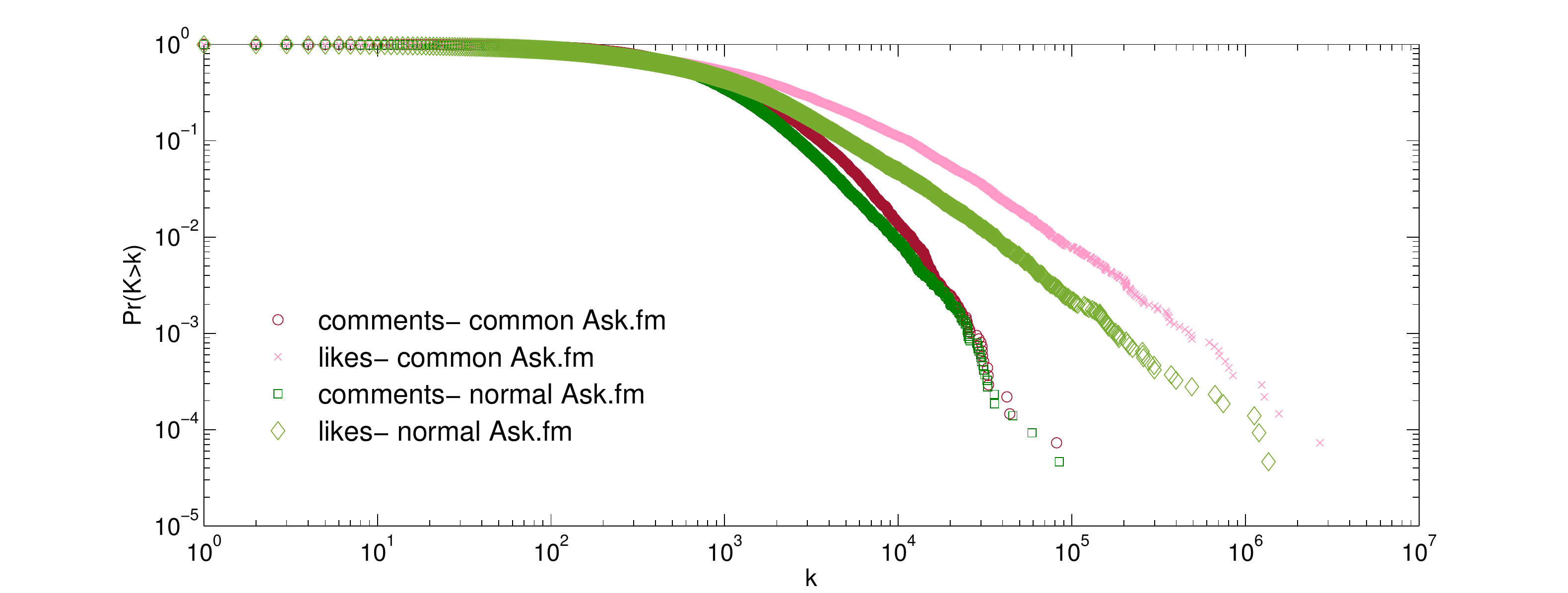}
\caption{CCDFs for the number of likes and comments on Ask.fm}
\label{ccdf_num_likes}
\end{figure}

We first analyze user activity in Ask.fm, which consists of the number of comments profile answer$+$question pairs and the total number of likes a user receives. For the rest of the paper we use the term comments for answer$+$question pairs which includes both comments by profile owners amd comments by other non-onwer users (loosley speaking, friends) for Ask.fm too.
Note that these two user activities are the only ones that we can
gather from the publicly accessible data of Ask.fm.  There are some
additional activities such as posting a question that is not answered by the
profile owner or user following activities that are not publicly available.
Figure \ref{ccdf_num_likes} shows the CCDFs of the two publicly available
user activities for normal and common users. 
We observe that common users are slightly more active in terms of received likes than normal users, 
but that otherwise the activity in terms of number of question+answer pairs is similar.
Therefore these common users exhibit opposite behavior in Ask.fm compared to Instagram relative to their normal CCDFs. It shows users who have two social network accounts (Ask.fm and Instagram) are more active/popular than the normal Ask.fm users and less active/popular than normal Instagram users.

\begin{figure}[!h]
\centering
\includegraphics[width=1\columnwidth]{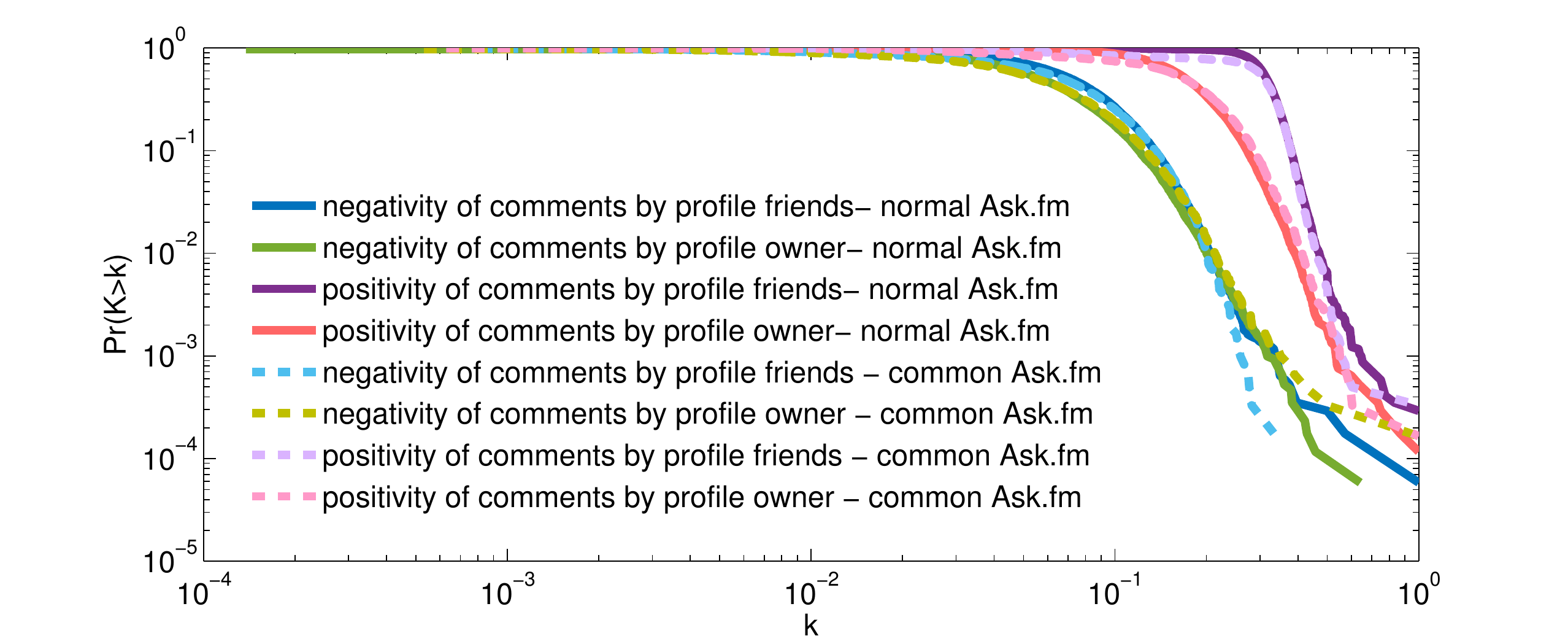}
\caption{CCDFs of the percentage of negative and positive posts by normal and common Ask.fm users. Red and purple show positivity and blue and green show negativity.}
\label{askfm_total}
\end{figure}

Finally, we analyze the user behavior in terms of negativity and positivity
in Ask.fm. As before, we divide all user posts into two
groups, those from the profile owner (answers) and those from friends (questions).
Figure \ref{askfm_total} shows the CCDFs of negativity and positivity of
comments for
normal and common users. In this figure, dashed lines are the CCDFs for
normal users and solid lines are for common users. 
As for Instagram, we observe on Ask.fm that positivity is always greater than 
negativity for both questions and answers for normal and common users.
\emph{As for Instagram, the second important observation we make is that the user behavior is
quite similar for normal and common users on Ask.fm both for negative and positive user
behaviors.}

\section{User Behavior Across Instagram and Ask.fm}
\label{sec:commn}

We now examine the impact of network features on user behavior. To do
so, we have identified a set of 8K users who are members of both
Ask.fm and Instagram social networks (common users). We will analyze the
activities of these users from their postings in the two networks. The
key question we seek to answer is ``How does their posting behavior differ
between the two social networks?''

\begin{figure*}[htbp!]
\centering
\includegraphics[width=2\columnwidth]{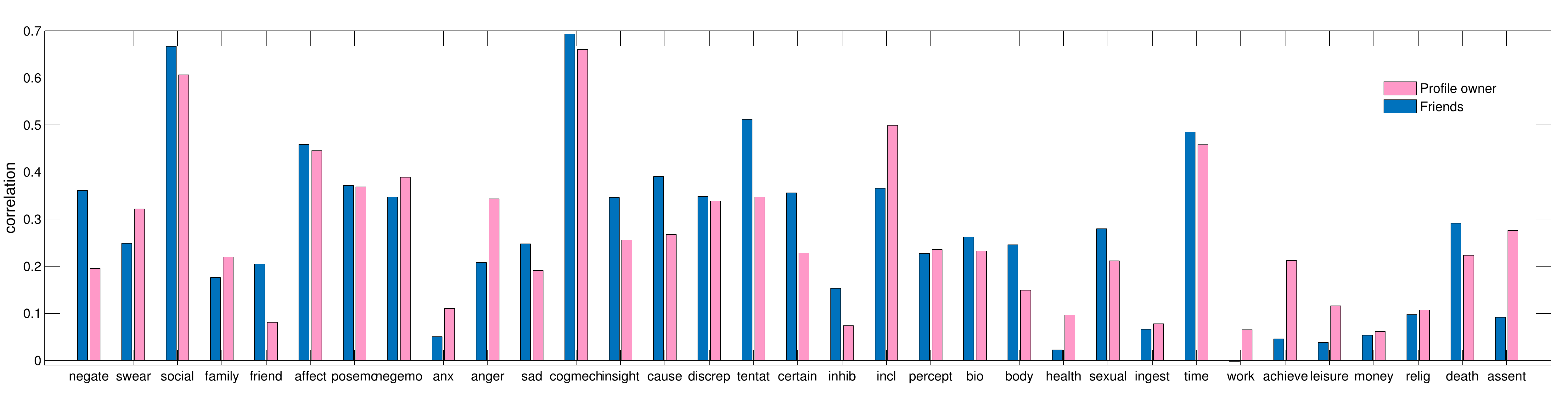}
\caption{Correlations of profile comments between two social networks for different liwc categories, separated by owner and friend comments ($p < 10^{-5}$).}
\label{liwc_crrltn}
\end{figure*}

We use Linguistic Inquiry and Word Count (LIWC), which is a text
analysis program that counts words and place messages in psychologically
meaningful categories \cite{liwc}. Four different sets of comments are considered for each user; comments by friends and comments by profile owners in Instagram and Ask.fm. We calculate the LIWC values for a subset of categories for these four types of comments for each common user.
In Figure \ref{liwc_crrltn}, we show the correlation for various categories
between the two different social networks for the postings by the profile
owners and the comments they receive from their friends. Some of these
categories have been chosen from \cite{TeenAsk} that introduces
some statistics about categories which teens talk about in their social
networks. In this figure, the correlation for the comments posted by a
profile owner across two networks is shown with pink bars, and the
correlation for the received comments from friends by a profile owner across
the two networks is shown in blue bars. 

We first note that there are certain categories such as social, cogmachin
and time for which the correlation is quite high for both posts by
profile owners and by friends. This indicates that the differences in the
two social networks have no impact in the discussion of these categories.
On the other hand, there are certain categories such as anx, inhib, health,
ingest, work and money for which the correlation is quite low for both
posts by profile owners and by friends. This indicates that the differences
in the two social networks have a significant impact in the discussion of
these categories.

Next we look at the differences in correlations between the comments by
profile owners and the comments by friends.
For categories like anxiety and anger which are personal characteristics,
the correlation between profile owner's comments across networks is higher
than correlation of friend's comments. For sad, sexual and death categories
the correlation is higher for friend's comments.  That is, if an owner has
sexual comments directed at them on one network, then there is a likelihood
that they will also have sexual comments directed at them on the second network.

Table \ref{stats} provides a summary of the statistics of the average number
of comments posted by owners or friends in Ask.fm and Instagram among
normal and common users. The average number of comments posted by friends
is larger than the comments by profile owner by a factor 3.4 for common
public Instagram users, and surprisingly by 16.3793 for public normal
users. Note that the ratio of owner to friend posts is 1 to 1 as expected for Ask.fm.
We further observe that while
common users have greater activity than normal users on Ask.fm, the
opposite is true on Instagram.  This agrees with our earlier analysis of the CCDFs
of Figures 2 and 4.

\begin{table*}[htbp!]
\centering
\begin{tabular}{|c|r|r|r|r|}
\hline 
        & Normal Instagram & Common Instagram & Normal Ask.fm & Common Ask.fm  \\ \hline
Avg \# posts by profile owner      & 195.90  & 116.06   & 1216.5 & 	 1496.50   \\ \hline
Avg \# posts by friends            & 3208.70 & 394.60     & 1216.5 & 	 1496.50  \\ \hline
Avg \% negative posts by profile owner & 3.72   &     3.39    &   6.54  &  6.47  	\\ \hline
Avg \% negative posts by friends       & 4.62      &   4.78      &  7.79  & 7.21	  	    \\ \hline
Avg \% positive posts by profile owner & 17.13  &   16.69      &  17.97   &  16.81     	\\ \hline
Avg \% positive posts by friends       & 27.59  &  26.4      &  29.18   & 	 27.53  	\\ \hline
\end{tabular}
\caption{Comparison of (positive and negative) posts by profile owners and friends across four different user groups.}
\label{stats}
\end{table*}

\begin{figure}[!h]
\centering
\includegraphics[width=1\columnwidth]{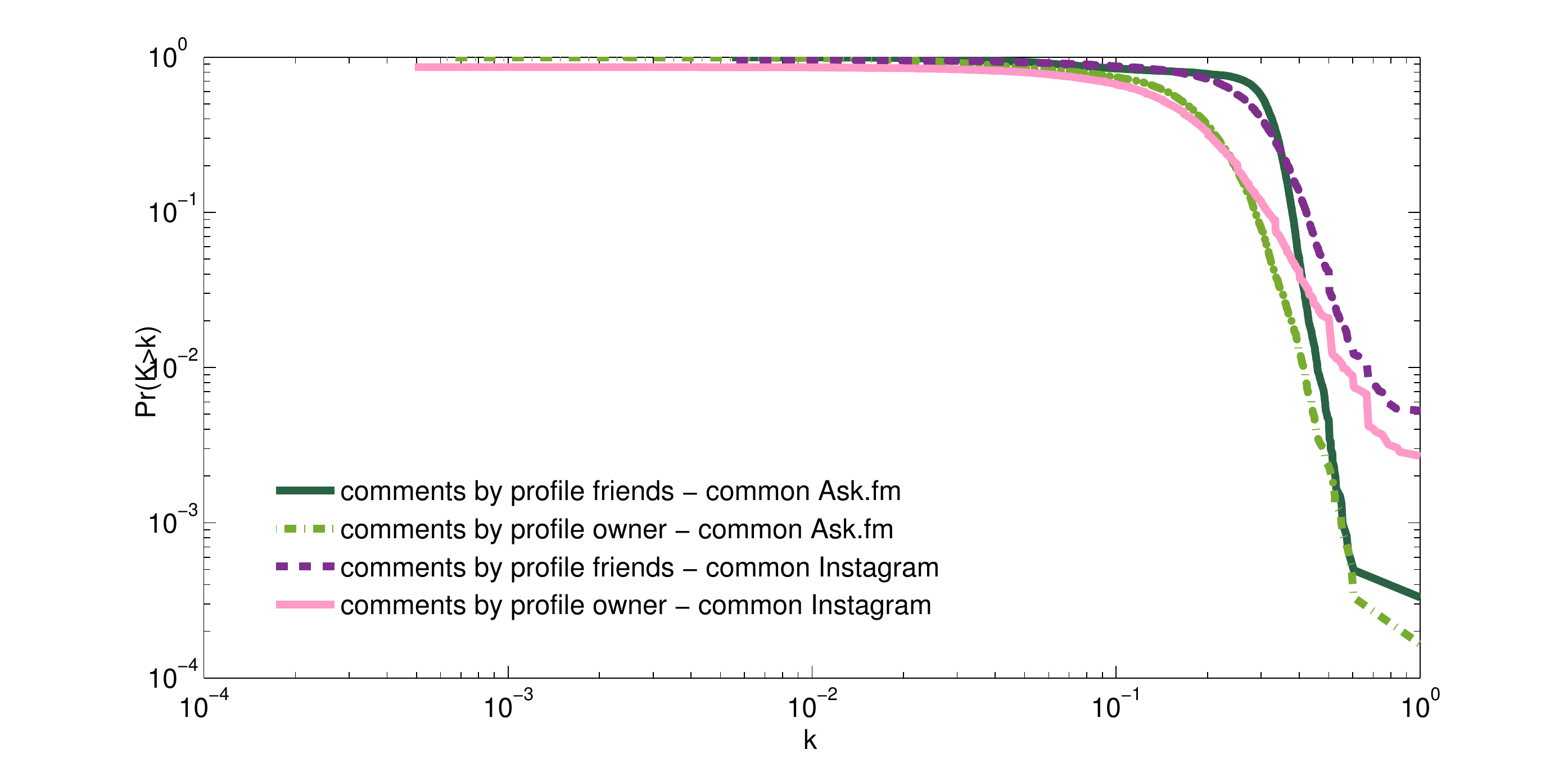}
\caption{CCDFs of the percentage of positive posts by common Instagram and Ask.fm users. Pink and purple show common Instagram users and light and dark green show common Ask.fm users.}
\label{comm_pos}
\end{figure}

\begin{figure}[!h]
\centering
\includegraphics[width=1\columnwidth]{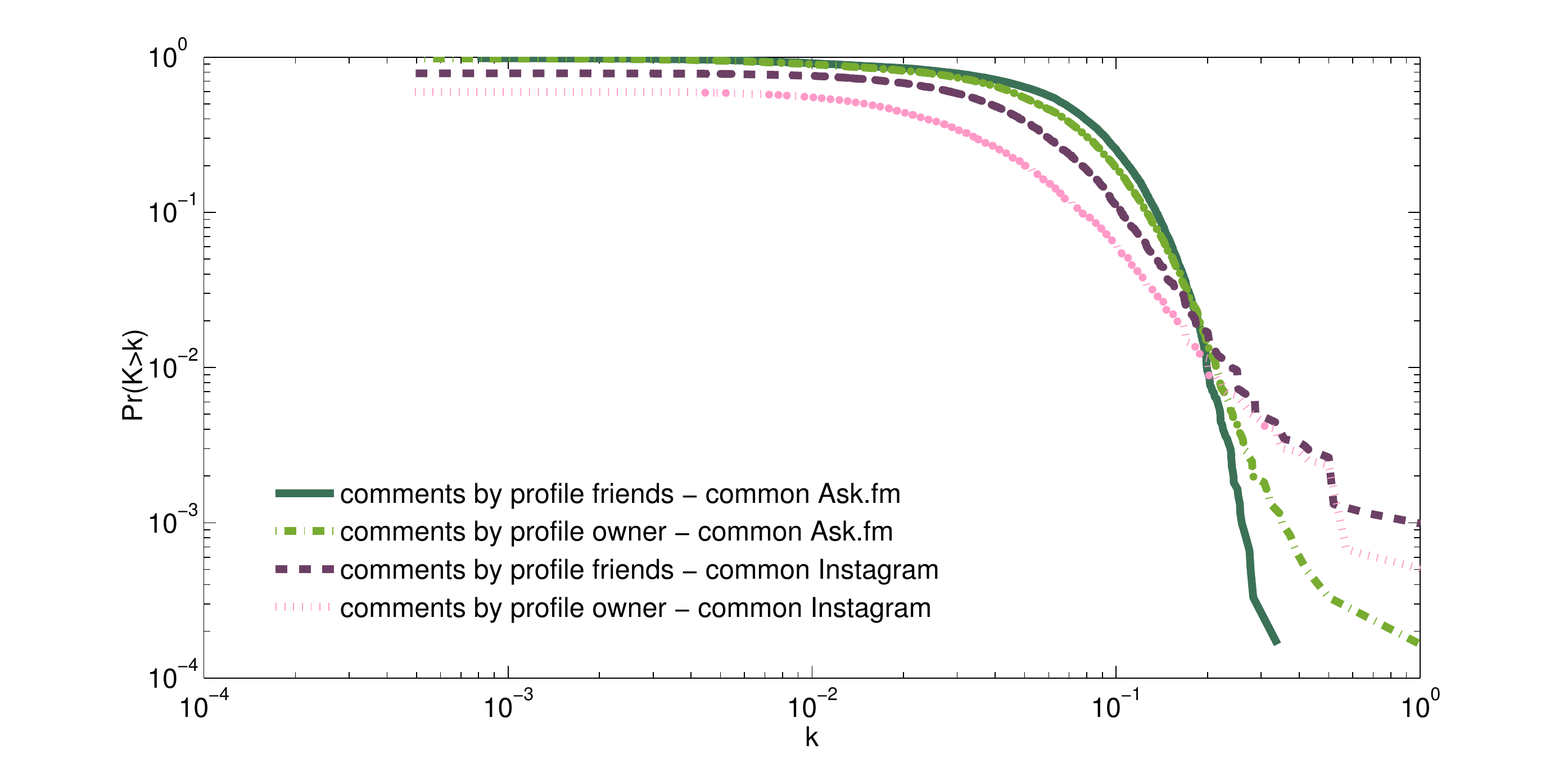}
\caption{CCDFs of the percentage of negative posts by common Instagram and Ask.fm users. Pink and purple show common Instagram users and light and dark green show common Ask.fm users.}
\label{comm_neg}
\end{figure}

To understand in more detail whether common users are more or less positive or negative
on Instagram vs Ask.fm, we calculated the percentage of positive posts and negative posts for each user profile.  We further divided the posts into those written by owners and friends.  Figures \ref{comm_pos} and \ref{comm_neg} show the CCDFs of positivity and
negativity respectively for the common users.  \emph{Figure \ref{comm_pos} shows that the positive
behavior of common users across both networks is fairly similar for both owners and friends.}
This is confirmed also by the similarity of average percentages of positivity shown in Table \ref{stats}.
The table also shows that positivity of friends' posts is consistently higher than owners' posts across all four categories in terms of average percentage of positive posts (by $t$-test with $p<0.0001$).


Figure \ref{comm_neg} compares the negativity of common users across the two networks.  \emph{We find among common users that most profile owners are more negative on Ask.fm than Instagram.  We also find that most friends are slightly more negative on Ask.fm than Instagram.}  This is confirmed from Table \ref{stats}, where we observe that the averages of the percentage of negative posts is higher on Ask.fm than Instagram, for both common and normal users (by $t$-test with $p<0.0001$).  Also, we note from the table that the negative percentages are clearly lower than the positive percentages across all 8 categories.  This confirms our earlier analysis for the CCDFs from Figures 3 and 5.


\begin{figure}[!h]
\centering
\includegraphics[width=1\columnwidth]{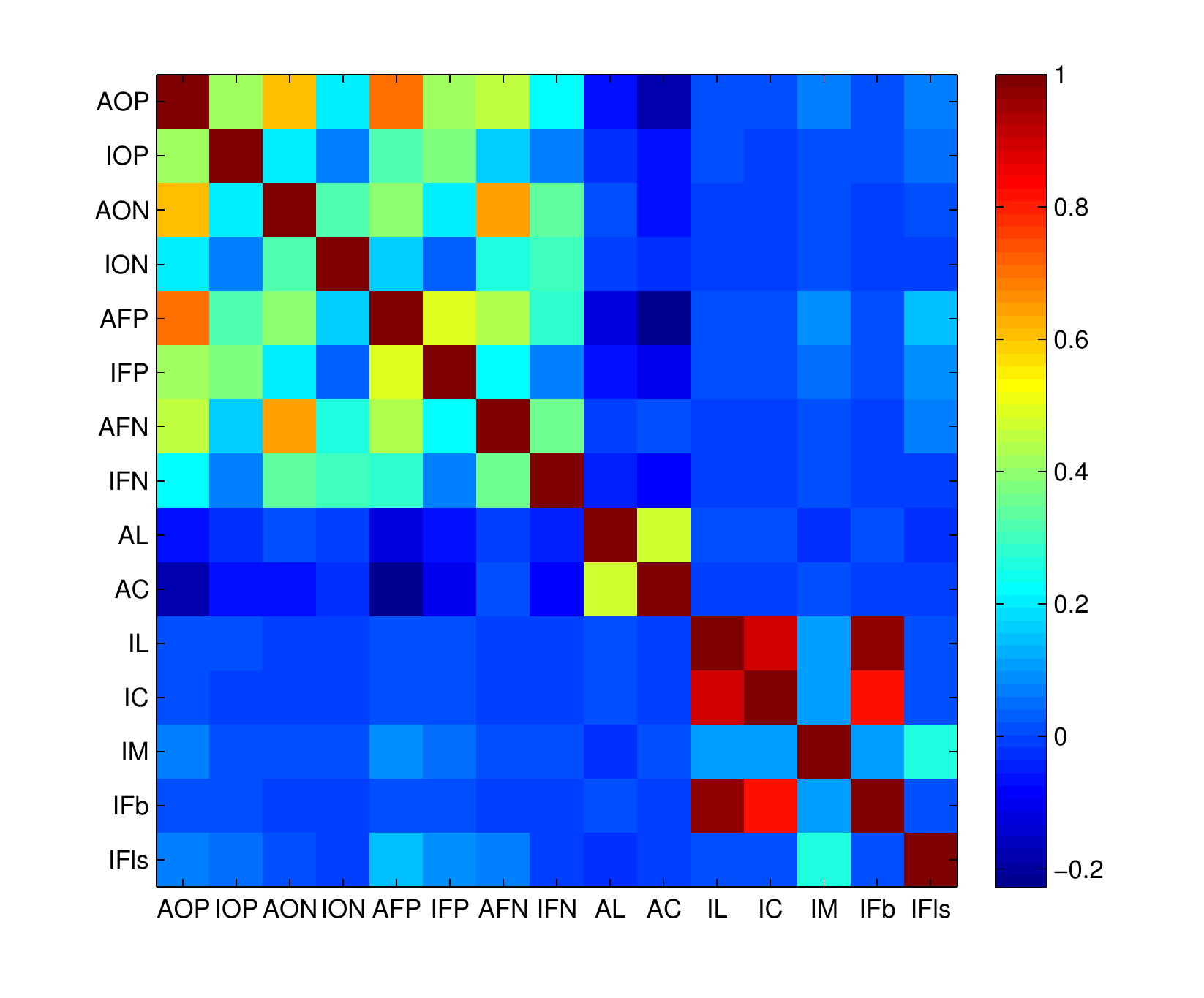}
\caption{Correlation matrix, where the characters stand for A: Ask.fm, I: Instagram, P: Positivity, N: Negativity, O: Owner's comment, F: Friend's comments, L: Likes, C: Comments, M: Media, Fb: Followed by, Fls: Follows ($p < 10^{-5}$).}
\label{crrltn_mtrx}
\end{figure}

Finally, we analyze the correlation among the positivity and negativity of
profiles across Ask.fm and Instagram and different social network activities or popularity factors (See Figure \ref{crrltn_mtrx}).
In the axes legends in this figure, the first (left side) eight x-axis
markers or the top eight y-axis markers represent positivity and negativity
in various users posts. The right seven markers on x-axis or the lower
seven markers on the y-axis represent various user activities.  Correlation
values only range from -0.2 to 1.0 because no lower values of negative
correlation were found.

First, we notice that there is no correlation in the upper right side and
lower left side in this figure. This means that there is no correlation
between the negativity or positivity percentage of profiles with any user
activities such as total comments, likes, followed by or follows in any
of the two social networks. 
Next, we notice that in the lower right side, there is significant
correlation with correlation values of 0.89 between Instagram likes
and comments (IL and IC), while the correlation between Ask.fm
likes and comments(AL and AC) is lower at 0.46.
We also find that the correlation
between the number of likes and posted media on Instagram (IL and IM) is only around
0.1. It seems that the users who receive more likes also have more comments
(sent+received), however the number of posted media does not seem to be a
factor for becoming more popular in terms of receiving more comments or
likes or being followed more. Also there is significant correlation,
correlation value 0.98, between the number of likes and followed by (IL and
IFb). Similarly, there is a high correlation between the number of
comments and followed by (IC and IFb). This is
not surprising as when you have more fans to follow you, you receive more
likes. 

Looking at the upper left side, the highest correlation, correlation value
0.7, is between positivity of Ask.fm owner's answers and friend's questions
(AOP and AFP).
Correlation between negativity of Ask.fm owner's answers and friend's questions is approximately the same with value 0.64 (AON and AFN). \emph{Considering the nature of the Ask.fm
social network where there is a specific answer for each question, this
means positive questions have been mostly replied with positive answers
and the negative questions are replied with negative answers.}

\begin{figure}[!ht]
\centering
\includegraphics[width=1\columnwidth]{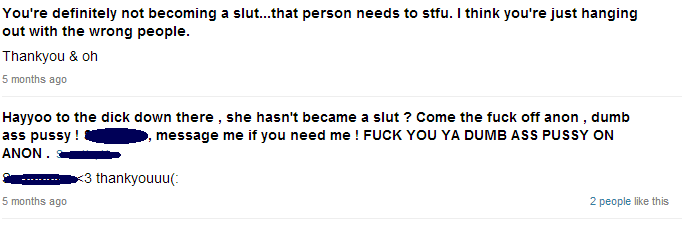}
\caption{An example of a profile in which negative words from a bystander which is not a cyberbullying question against profile owner.}
\label{bystanders}
\end{figure}

Also, the correlation between answers' positivity and questions'
negativity (AOP and AFN) in Ask.fm is around 0.45. Based on our observation
of some user profiles at Ask.fm, we noted that a lot of negative
questions actually come from user's friends. Some come from upstanders
when a cyberbullying case happens for the profile owner and some are
just friendly talk between close friends (See Figure \ref{bystanders}).  

On the other hand, we can not see such a correlation between comments
by profile owners and comments by friends in Instagram. The reason for this
is the fact that in Instagram, one answer can be used in reply to several
received comments, e.g. ``Thank you all'' is just one answer to several
greeting messages received. The correlation between Instagram received
and sent comments is just around 0.3 for both negativity and positivity,
which indicates that there is less correlation between Instagram comments.

Another interesting observation is the correlation of 0.4 between
positive behavior of profile owners across their Ask.fm and Instagram
(AOP and IOP).  The correlation of negative behavior of profile owners across
the two networks (AON and ION) is a approximatley similar value.
However, positive behavior of a profile owner in one network is not 
correlated with the negative behavior of the profile owner in the other
network (AOP and ION or AON and IOP). \emph{These results provide some support
for the notion that a profile owner who is positive on one network generally tends to be positive
on the other network, and similarly, a profile owner who is negative on
one network generally tends to be negative on the other network.}

\section{Impact of Anonymity on User Behavior in Ask.fm}

One important feature in a social network that can potentially lead to negative
user behavior is anonymity. Intuitively, it seems that if users can post 
messages anonymously, they will tend to be more negative. Indeed, anecdotal
evidence suggests this. For example, according to Natasha MacBryde's father,
anonymity had a big role in his daughter's suicide \cite{Bullies}.
To understand the impact of anonymity on user behavior, we analyzed the data
we collected from Ask.fm. Recall that Ask.fm is a semi-anonymous network in
which users asking questions are anonymous by default and some users
choose to reveal themselves while asking questions. So, we analyzed 
negativity in questions that were posted anonymously and compared that with
negativity in questions in which users chose to reveal themselves.

\begin{figure}[!h]
\centering
\includegraphics[width=1\columnwidth]{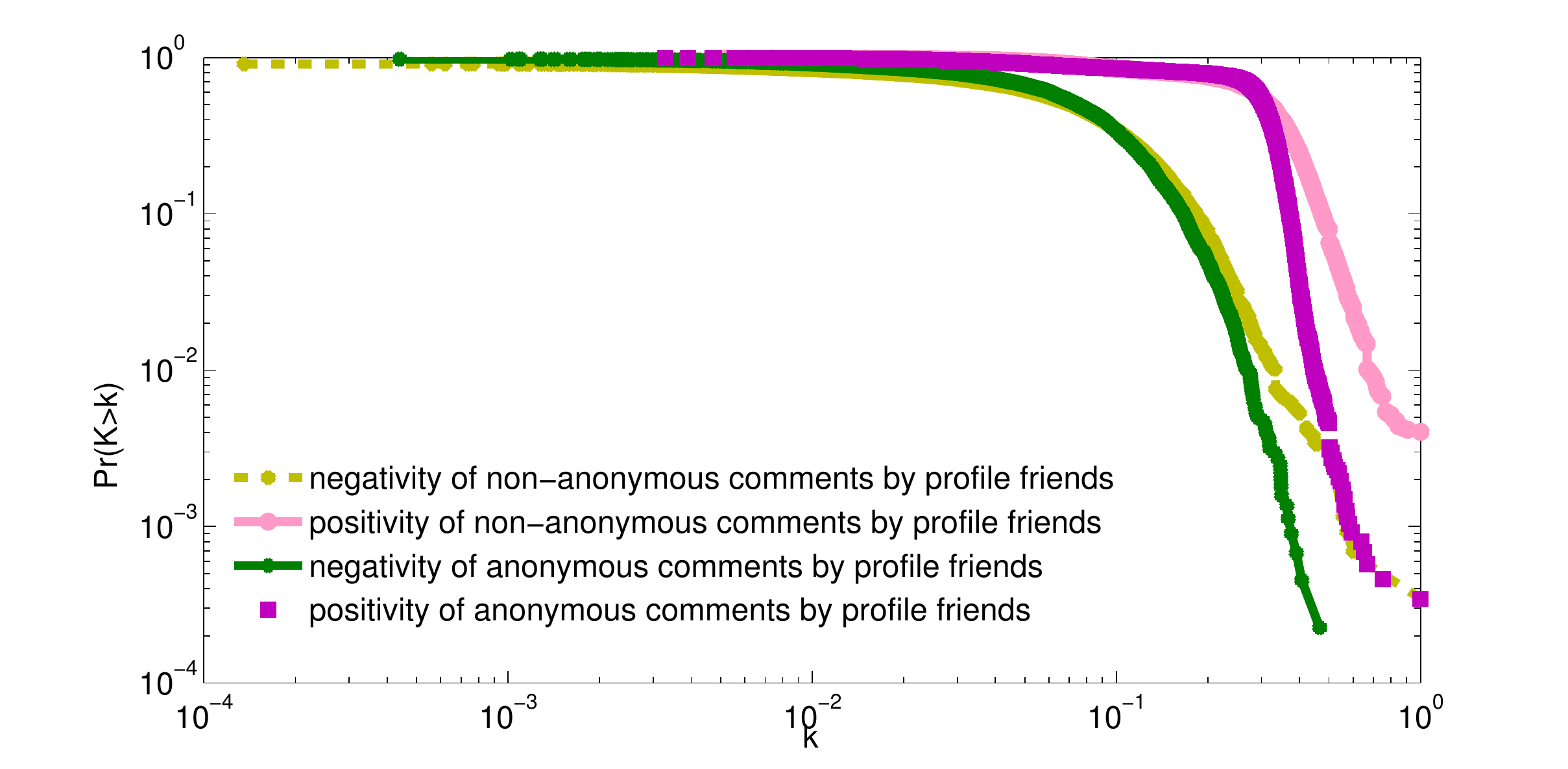}
\caption{CCDFs of the percentage of positive and negative posts for Ask.fm users. Pink and purple show positivity and light and dark green show negativity.}
\label{anon}
\end{figure}

In particular, we collected the profile contents of about 15K Ask.fm users (10.6K normal and 4.4K common users)
Figure \ref{anon} shows the CCDFs of the percentage of positive and negative
questions for Ask.fm users. \emph{We notice that the the non-anonymous comments tend
to be more negative and more positive than the
anonymous comments.} This is a surprising result, since we expected that 
non-anonymous users will be less negative. To gain a better insight into
these behaviors, we delved into some of the users' profile contents and
found that the reasons of non-anonymous comments having more negativity is
twofold. First, some non-anonymous comments are from users who try to defend
a profile owner from abusive comments that were posted anonymously on his/her
profile. In turn, these supporting comments from non-anonymous users end
up containing
some negative words. For example, a non-anonymous comment in one of the
user's profile whose profile content has several negative anonymous comments
says \textit{``Hayyoo to the d**k down there , she hasn't became a s**t ? Come the f**k off anon , dumb a** p**sy ! Schuyler , message me if you need me ! F**K YOU YA DUMB A** P*SSY ON ANON''} as it is shown in Figure \ref{bystanders}. In this comment, the non-anonymous user attacks anon (anonymous user) for abusing the profile owner. Another example of such a post goes like this, \textit{``All you people calling her a s**t, GET OVER YOUR SELF! Your probably the ones who are the s**ts and are fake! Just leave her alone b*tchez! (*****, they are all just jealous of you!)''}.

\begin{figure}[!h]
\centering
\includegraphics[width=0.95\columnwidth]{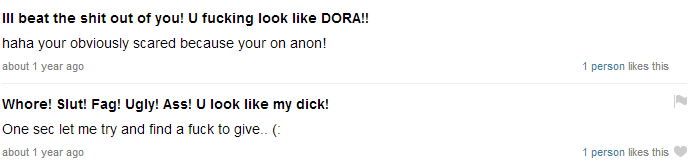}
\caption{An example of a profile having posts with negative words from an anonymous user which is cyberbullying question against profile owner }
\label{ano_abuser}
\end{figure}

\begin{figure}[!h]
\centering
\includegraphics[width=0.95\columnwidth]{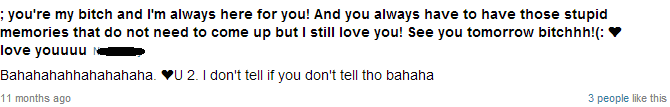}
\caption{An example of a profile having posts with affectionate negative words from a non-anonymous user}
\label{affectionate_negative}
\end{figure}

The second reason for increased negativity in questions posted by
non-anonymous users is that  some non-anonymous comments are from close
acquaintances who use negative words as a mean for showing affection.
For example, a non-anonymous comment goes like this/ \textit{``you're my b**ch
and I'm always here for you! And you always have to have those stupid
memories that do not need to come up but I still love you! See you tomorrow
bi***hh!(: love youuuu''} (See Figure \ref{affectionate_negative}). On the
other hand, there exist negative anonymous comments that tend to be actually abusive. For
example, \textit{``Wh**e! S**t! Fag! Ugly! A*s! U look like my d**k!''}
or \textit{``yhu trynna s**k for a buck b**ch''} as shown in Figure
\ref{ano_abuser}. So, it is evident from the above discussion and examples
that negativity in a comment is not always an indicator of cyberbullying, and that
more research, such as labeling comments as cyberbullying or not, will be needed
to firmly establish the relationship between anonymity and cyberbullying.


\section{Discussion}

We found that the users are more negative in Ask.fm than Instagram.   It is tempting to attribute the increased negativity on Ask.fm to the anonymity of Ask.fm.  However, there are other differences between the two networks that could influence the increased negativity on Ask.fm.  For example, Ask.fm users primarily post in text, whereas Instagram is highly focused on posting image media, followed by the commenting.  Also, the relationship between Ask.fm owners comments and non-owner ``friends" comments is one-to-one, whereas on Instagram, a single owner comment may be directed at multiple friends' comments, such as a "Thank you all".  This requires more research into the relationship between anonymity and negativity. 

While we have conducted a negative (and positive) word analysis on posting, we would like to extend this research.  Not all occurrences of negative words correspond to cyberbullying, e.g. ``that's f**king amazing".  Towards this end, we would need to specifically label comments as being examples of cyberbullying.  

We would like to extend our negativity analysis to understand the degree of negativity, and how the targeted owners are affected by the degree of negativity.  Highly negative profiles with a high percentage of negative posts with little positive support from others may reveal the most vulnerable victims to cyberbullying.  Also, while we have explored anonymity and its association with negativity within a single social network of Ask.fm, we wish to extend this work to other social networks.



\section{Conclusions}


We have studied user behavior in two social networks, Instagram and Ask.fm, that are among the most popular used for cyberbullying.  We have analyzed negative and positive words usage from comments by users common to both the Instagram mobile social network and the Ask.fm online social network.  The following comprise our key findings.  First, we found that both owners and ``friends" who post on common users'  profiles have increased negativity on Ask.fm compared to Instagram.  Second, there is no difference in the positivity of both owners and friends on each network.  Third, common users are similar in their positivity or negativity when compared to normal non-common users on each of the two networks.  Fourth, anonymity actually results in less negativity on Ask.fm compared to non-anonymous comments.  We have described where more research is needed to label which negative comments constitute cyberbullying.  Finally,  we found that there is a strong correlation on Ask.fm between owners and friends' negative comments, and similarly for positive comments.  There is also some support for the notion that a profile owner who is positive on one network generally tends to be positive on the other network, and similarly for negative behavior.


\section{Acknowledgements}

We would like to thank the US National Science Foundation (NSF) funding agency for supporting this research through grant number 1162614.




%
\IEEEpeerreviewmaketitle

\bibliographystyle{IEEEtran}
\bibliography{ref}
\end{document}